\documentclass[conference]{IEEEtran}
\ifCLASSINFOpdf
\else
 \usepackage[dvips]{graphicx}
 \graphicspath{{./figures/}}
\fi
\usepackage{amsmath}
\usepackage{balance}

\usepackage{algorithmic}
\ifCLASSOPTIONcompsoc
\usepackage[caption=false,font=normalsize,labelfont=sf,textfont=sf]{subfig}
\else
\usepackage[caption=false,font=footnotesize]{subfig}
\fi
\usepackage{url}
\usepackage{amsfonts}
\usepackage{amssymb}
\usepackage{amsthm}
\usepackage{commath}

\DeclareMathOperator*{\argmin}{arg\,min}
\newtheorem{theorem}{Theorem}

\newtheorem{proposition}[theorem]{Proposition}

\usepackage{algorithm}

\begin{document}
	\title{Kernel Node Embeddings}

	\author{\IEEEauthorblockN{Abdulkadir \c{C}elikkanat}
		\IEEEauthorblockA{CentraleSup\'{e}lec and Inria Saclay\\
			University of Paris-Saclay\\
			Gif-Sur-Yvette, France\\
			Email: abdulkadir.celikkanat@centralesupelec.fr}
		\and
		\IEEEauthorblockN{Fragkiskos D. Malliaros}
		\IEEEauthorblockA{CentraleSup\'{e}lec and Inria Saclay\\
			University of Paris-Saclay\\
			Gif-Sur-Yvette, France\\
			Email: fragkiskos.malliaros@centralesupelec.fr}}
	
	
	%Telephone: (800) 555--1212, Fax: (888) 555--1212}


	\maketitle
	
	\begin{abstract}
		Learning representations of nodes in a low dimensional  space is a crucial task with many interesting applications in network analysis, including link prediction and node classification. Two popular  approaches for this problem include \textit{matrix factorization} and \textit{random walk}-based models. In this paper, we aim to bring together the best of both worlds, towards learning latent node representations. In particular, we propose a weighted matrix factorization model which encodes random walk-based information about the nodes of the graph. The main benefit of this formulation is that it allows to utilize  kernel functions on the computation of the embeddings. We perform an empirical evaluation on real-world networks, showing that the proposed model outperforms baseline node embedding algorithms in two downstream machine learning tasks.
	\end{abstract}
	
	% Although there is a variety of methods, the random walk and matrix factorization based approaches are two prominent classes and  in this paper we bring them together with the kernel functions. The kernels are generally very well-known in the context of machine learning algorithms such as \textit{Support Vector Machines} and \textit{Principal Component Analysis} but we consider them in learning node embeddings by generalizing the existing assumptions about the interactions between the pair of nodes.
	

	% \fi
	\IEEEpeerreviewmaketitle

	\section{Introduction}\label{sec:introduction}
	% With the advancements in technology, the process of storing and recording the information has become easier and the size of the collected data has been continuously increasing. This situation requires more tools for processing and analyzing this huge amount of information and the graph structures are just one of these helpful abstract elements which are commonly used in various disciplines ranging from physics to biology in order to represent and to model the highly complicated relationships among different entities. Therefore, in recent years, the network analysis has gained considerable interest to extract meaningful and consistent information from networks and many approaches have been developed to grasp their inherent characteristics.
	
	With the advancements in data production, storage and consumption, networks are becoming omnipresent; data from diverse disciplines can be represented as graph structures with prominent examples here being various social, information, technological and biological networks. Developing machine learning algorithms to analyze, predict and make sense of the structure of graph data has become a crucial task with a plethora of cross-disciplinary applications \cite{survey_hamilton_leskovec, survey}. 
	%Unlike the classical tools directly utilizing the structure of networks \cite{laplacian_eigenmap, grarep, hope, adaptive_diffusion}, a recent prominent strategy, named by \textit{network representation learning (NRL)}, aims at finding vectors corresponding to nodes in such a way that some properties of networks are intended to be captured by means of the relative positions of embedding vectors in a latent space with respect to each other \cite{deepwalk, node2vec, biasedwalk}. These embedding vectors are later used in performing various downstream tasks \cite{survey_hamilton_leskovec, node_embed_comm}.
	The major challenge in machine learning on graph data concerns the
	encoding of information about its structural properties into the learning model. To this direction, a recent paradigm in network analysis, known as \textit{network representation learning} (NRL), aims at embedding the nodes of the graph into a lower-dimensional space, in such a way that similarity among nodes in the graph  is captured by the similarity of the embeddings in the latent space \cite{deepwalk, node2vec, laplacian_eigenmap, grarep, hope, adaptive_diffusion}. Many of the proposed models in network representation learning have mostly concentrated on computing node embeddings relying on matrix factorization techniques that encode information about structural node similarity \cite{laplacian_eigenmap, grarep, hope}. Nevertheless, the majority of those approaches are not efficient for large scale networks, mainly due to the high computational cost required to perform matrix factorization \cite{survey_hamilton_leskovec, survey}.
	
	\par Being inspired by the field of natural language processing \cite{word2vec},  random-walk based models have gained considerable attention \cite{deepwalk,node2vec,biasedwalk, tne}. Typically, these methods first generate a set of node sequences (i.e., \textit{context} nodes) for every node (i.e., \textit{center}) in the network, based on some random walk strategy; then, node representations are learned by predicting context-center node co-occurrences within the random walks.
	%Although the random walk based approaches are very similar in modeling the relationships between nodes of the network, the strategy that they follow to perform random walks is the key particularity that makes them distinct from each other \cite{deepwalk, node2vec, biasedwalk}.
	
	In this paper, we aim at combining the previously proposed broad modeling approaches for NRL -- namely matrix factorization and random walks. In particular, we focus on modeling the interactions between nodes based on random walks, under a weighted matrix factorization framework. The potential advantage of such a modeling approach is that it allows to take advantage of and combine the elegant mathematical formulation that matrix factorization can offer with the expressive power of random walks to capture a notion of ``stochastic'' node similarity in an efficient way. More importantly, this formulation allows us to utilize \textit{kernel} functions in the node representation learning task.
	
	Kernel functions have mostly been introduced along with popular  learning algorithms, such as \textit{PCA} \cite{kernel_pca}, \textit{SVMs} \cite{svm}, \textit{Spectral Clustering} \cite{kernel_clustering} and \textit{Collaborative Filtering} \cite{kernel_matrix_fact}. The idea is to map non-linearly separable points into a (generally) higher dimensional feature space, so that the inner product in the new space can be computed without needing to compute the exact feature maps.
	%We also consider the kernels for a similar purpose but conversely, our purpose is to obtain data points by given values representing the relationships among nodes.
	Here, we aim at obtaining embeddings, given values that represent the relationships among nodes.  Because of the nature of matrix factorization-based methods, these values are viewed as an inner product of vectors lying on a latent space, which allows us to utilize kernels interpreting the embeddings in a higher dimensional feature space using non-linear maps. The main contributions of the paper are the following:
	\begin{itemize}
		\item We propose a novel approach for learning node embeddings by incorporating kernel functions with models relying on weighted matrix factorization, encoding random walk-based structural information of the graph.
		\item We extensively evaluate the performance of the proposed method in the downstream tasks of node classification and link prediction and we show that the model generally outperforms the well-known baseline methods on various network datasets.
	\end{itemize}
	
	\noindent \textbf{Notation.} We use the notation $\mathbf{M}$ to denote a matrix, $\mathbf{M}_{i,j}$ points out the entry located at the $i$'th row and $j$'th column of the matrix, and $\mathbf{M}_{i,:}$ indicates the $i$'th row of the matrix.
	%For a matrix $\mathbf{M}$, we use $\mathbf{M}_{i.j}$ to denote the entry at the $i$'th row and $j$'th column and  $\mathbf{M}_{(i.:)}$ shows the $i$'th row. The \textit{Frobenius} norm of the matrix is denoted by $\Vert \mathbf{M} \Vert_{F}$.
	%which is defined by $\sqrt{\sum_{i,j>0}\mathbf{M}_{i,j}^2}$
	%Kadir, we might add 2-3 lines here about the notation -- not the symbols but the convention that we follow for matrices, etc. Also, norms, etc.
	
	\vspace{.2cm}
	
	\noindent \textbf{Source code.} The C++ implementation of the proposed methodology and the networks used in the study, can be reached at: \url{https://abdcelikkanat.github.io/projects/kernelNE/}.

	\section{Modeling and Problem Formulation}
	
	% \begin{table}[t]
	%     \centering
	%     \caption{This table is only for keeping track of the symbols and will be removed}
	%     \label{tbl:symbols}
	%     \begin{tabular}{cl} 
	%     \hline
	%     \textbf{Symbol}     &  \textbf{Description} \\
	%     \hline
	%     $G$ & Graph \\
	%     $\mathcal{V}$ & Vertex set \\
	%     $\mathcal{E}$ & Edge set \\
	%     $d$ & Dimension of embedding vectors \\
	%     $\alpha$, $\beta$     & Latent representations of nodes \\
	%     $\mathcal{C}$ & The derived matrix \\
	%     $\boldsymbol{w}$ & A sequence of nodes, a random walk \\
	%     $\gamma$ & Window size\\
	%     $\mathcal{W}$ & A sequence of walks \\
	%     $\mathcal{N}_{\gamma}(v)$ & The context sequence of node $v$\\
	%     $\mathcal{C}_{\gamma}(v)$ & The context set of node $v$\\
	%     $N$ & The number of walks\\
	%     $L$ & The length of walks\\
	%     $\eta$ & Natural parameter of an exponential family\\
	
	%     $f$ & Link function\\
	%     \hline
	%     \end{tabular}
	% \end{table}
	
	Let $G=(\mathcal{V}, \mathcal{E})$ be a graph where $\mathcal{V}=\{1,...,n\}$ and $\mathcal{E}\subseteq\mathcal{V}\times\mathcal{V}$ are the vertex and edge sets, respectively. Our goal is to find node representations in a latent space, preserving properties of the network. More formally, we define the general objective function  of our problem as a weighted matrix factorization \cite{weighted_low_rank}, as follows:
	
	\begin{align}\label{eq:main_obj_func}
	\argmin_{\mathbf{A},\mathbf{B}}\frac{1}{2}\left\Vert\mathbf{W}\odot(\mathbf{M} - 
	\mathbf{A}\mathbf{B}^{\top})\right\Vert_F^2,
	\end{align}
	
	\noindent where $\mathbf{M}\in\mathbb{R}^{n\times n}$ is the target matrix constructed based on the desired properties of a given network, which is used to learn node embeddings $\mathbf{A}, \mathbf{B}\in \mathbb{R}^{n \times d}$. $\mathbf{W} \in \mathbb{R}^{n \times n}$ is the weight matrix in which each element $\mathbf{W}_{v,u}$ captures the importance of the approximation error between nodes $v$ and $u$, and $\odot$ indicates the \textit{Hadamard} product. Depending on the desired graph properties that we are interested to encode, there are many possible alternatives to choose matrix $\mathbf{M}$; such include the number of common neighbors between a pair of nodes, higher-order node proximity based on the \textit{Adamic-Adar} or \textit{Katz}  indices \cite{hope}, as well based on $k$-hop information \cite{grarep}. Here, we will design $\mathbf{M}$ as a sparse binary matrix utilizing information of random walks over the network. Note that,  matrices $\mathbf{M}$ and $\mathbf{W}$ do not need to be symmetric.
	
	%Although there are numerous possible choices for the matrix $\mathbf{M}$, we consider it as a binary matrix where each entry $M_{v,u}\in\{0,1\}$ indicates the status of the node $u$ with respect to $v$ and $\mathbf{W}_{v,u}$ denotes the strength of the relationship. Note that the matrices does not have to be symmetric and they will be constructed by using random walks.
	
	Random walk-based node embedding models \cite{deepwalk, node2vec, biasedwalk, epasto-splitter, NguyenLRAKK18, node_embed_comm} have received great attention because of their good prediction performance and efficiency on large scale networks. Typically, those models generate a set of node sequences by simulating random walks; node representations are then learned by optimizing a model which defines the relationships between nodes and their \textit{contexts} within the walks. More formally, for a random walk  $\boldsymbol{w}=(w_1,...,w_{\ell})$, the context of the \textit{center} node $w_l\in\mathcal{V}$ at position $l$ in the walk $\boldsymbol{w}$ is defined as $\mathcal{C}_{\boldsymbol{w}}(w_l) := (w_{l-\gamma},...,w_{l-1},w_{l+1},...,w_{l+\gamma})$,  where $\gamma$ is called the \textit{window size} and it denotes the furthest distance between the \textit{center} and \textit{context} nodes $w_{k}\in\mathcal{V}$ for $l-\gamma \leq k \leq l+\gamma$ and $k\not=l$. The embedding vectors are then obtained by maximizing the likelihood of occurrences of nodes within the context of given center nodes. Here, we will also follow a similar random walk strategy, formulating the problem under a matrix factorization framework.
	
	Let $\mathbf{M}_{v,u}$ be a binary value representing if  node $u$ appears in the context of $v$ in any walk. Also, let $\mathbf{F}_{v,u}$ be the number of occurrences of  node $u$ in the contexts of $v$ in the generated walks. Setting each term $\mathbf{W}_{v,u}$ as the square root of $\mathbf{F}_{v,u}$, the objective function in (\ref{eq:main_obj_func}) can be expressed under a random walk-based formulation as follows:
	
	\begin{align}
	&\argmin_{\mathbf{A},\mathbf{B}}\frac{1}{2}\Big\Vert\sqrt{\mathbf{F}}\odot \big(\mathbf{M} - \mathbf{A}\mathbf{B}^{\top} \big)\Big\Vert_{F}^2\nonumber \nonumber\\
	=&\argmin_{\mathbf{A},\mathbf{B}}\frac{1}{2}\sum_{v \in \mathcal{V}}\sum_{u \in \mathcal{V}}\mathbf{F}_{v,u} \Big(\mathbf{M}_{v,u}  - \langle \mathbf{A}_{v,:},\mathbf{B}_{u,:}\rangle \Big)^2\nonumber\\
	=&\argmin_{\mathbf{A},\mathbf{B}}\frac{1}{2}\sum_{\boldsymbol{w}\in\mathcal{W}}\sum_{w_l \in \boldsymbol{w}}\sum_{u \in \mathcal{V}}\Big(\mathbf{M}_{w_l,u}^{\textbf{w}} - \langle \mathbf{A}_{w_l,:},\mathbf{B}_{u,:}\rangle \Big)^2\label{eq:objective_without_kernel},
	\end{align}
	
	\noindent where each $\boldsymbol{w}\in\mathcal{V}^{\ell}$ indicates a random walk of length $\ell$ in the collection $\mathcal{W}$ and $\mathbf{M}_{w_l,u}^{\textbf{w}}$ represents the occurrence of $u$ in the context $\mathcal{C}_{\textbf{w}}(w_l)$. Matrix $\mathbf{A}$ in Eq. \eqref{eq:objective_without_kernel},  contains the embedding vectors of nodes when they are considered as \textit{centers}; those will be the embeddings that are used in the experimental evaluation. The choice of matrix $\mathbf{M}$ and the reformulation of the  objective function as stated above, offers a computational advantage during the optimization step. More importantly, as we will present in the next section, we can further benefit from a \textit{kernelized} version of the objective function. 
	%the rows of the matrix $\mathbf{B}$ represent the embeddings when they are interpreted as \textit{context} nodes. 
	
	\section{Kernel-based Representation Learning} \label{sec:kernel}
	%In this section, we present our method by unifying the objective functions stated for random walks and matrix factorization approaches but we will firstly describe some fundamental concepts for unfamiliar readers. 
	
	% There are several methods \cite{netsmf-www2019, line} using matrix factorization techniques to find node representation in a lower dimensional space $(d \ll n)$ and the most of them adopts \textit{Singular Value Decomposition (SVD)} since it proposes the best approximation for the objective function in (\ref{eq:main_obj_func}) with a uniform weight matrix \cite{svd}.
	
	% Although other MF techniques apply SVD ...
	
	Similar to other matrix factorization techniques that aim at finding latent representations in a lower dimensional space $(d \ll n)$ (e.g., \cite{netsmf-www2019, line, netmf}), one can adopt \textit{Singular Value Decomposition} (SVD) to provide the best approximation of the objective function in (\ref{eq:main_obj_func}),  as long as the weight matrix is uniform \cite{svd}. It is also implicitly assumed that every element of the target matrix $\mathbf{M}$ can be written as inner product of vectors in the latent space, and in that case, it  becomes difficult to obtain an exact low-rank decomposition. To overcome this limitation, in our approach we utilize kernel functions to learn node representations via matrix factorization. 
	
	Let $(\mathsf{X}, d_X)$ be a metric space and $\mathbb{H}$ be a Hilbert space of real-valued functions defined on $\mathsf{X}$. A Hilbert space is called  \textit{reproducing kernel Hilbert space (RKHS)} if the point evaluation map over $\mathbb{H}$ is a continuous linear functional. Furthermore, a \textit{feature map} is defined as a function $\Phi:\mathsf{X}\rightarrow\mathbb{H}$, $\mathbb{H}$ is referred to as \textit{feature space} and every feature map defines a \textit{kernel} $\kappa:\mathsf{X}\times\mathsf{X} \rightarrow \mathbb{R}$ as follows:
	\begin{align*}
	\kappa(x,y) := \langle \Phi(x), \Phi(y) \rangle && \forall (x,y) \in \mathsf{X}^2.
	\end{align*}
	It can be seen that $\kappa(\cdot,\cdot)$ is symmetric and positive definite due to the properties of an inner product space.
	
	A function $g:X\rightarrow\mathbb{R}$ is called \textit{induced by $\kappa$}, if there exists $h\in\mathbb{H}$ such that $g=\langle h,\Phi(\cdot) \rangle$, for a feature vector $\Phi$ of kernel $\kappa$ (note that,  the definition is independent of the feature map $\Phi$ and space $\mathbb{H}$) \cite{kernel_steinwart}. Let $\mathcal{I}_{\kappa} := \{g:\mathsf{X}\rightarrow\mathbb{R} \ | \ \exists h\in\mathbb{H} \ \text{s.t.} \ g=\langle h,\Phi(\cdot)\rangle\}$ be the set of induced functions by kernel $\kappa$. Then, a continuous kernel $\kappa$ on a compact metric space $(\mathsf{X},d_{\mathsf{X}})$ is called \textit{universal}, if the set $\mathcal{I}_{\kappa}$ is dense in $\mathcal{C}(\mathsf{X})$. In other words, for any function $f \in \mathcal{C}(\mathsf{X})$ and $\epsilon > 0$, there exists $g_h \in \mathcal{I}_{\kappa}$ satisfying
	\begin{align*}
	\left\Vert f-g_{h} \right\Vert_{\infty} \leq \epsilon,
	\end{align*}
	where $g_h$ is defined as $\langle h, \Phi(\cdot) \rangle$ for some $h \in \mathbb{H}$. We use the next proposition as the basis for our approach.
	
	\begin{proposition}[\cite{kernel_steinwart}]
		Let $(\mathsf{X}, d)$ be a compact metric space and $\kappa(\cdot, \cdot)$ be a universal kernel on $\mathsf{X}$. Then, for all compact and mutually disjoint subsets $\mathsf{K}_1,...,\mathsf{K}_n \subset \mathsf{X}$, all $\alpha_1$,...,$\alpha_n$ $\in \mathbb{R}$, and all $\epsilon > 0$,  there exists a function $g$ induced by $\kappa$ with $\norm{g}_{\infty} \leq \max_i|\alpha_i| + \epsilon$ such that
		\begin{align*}
		\norm{g_{|K} - \sum_{i=1}^{n}\alpha_i\boldsymbol{1}_{K_i}}_{\infty} \leq \epsilon, 
		\end{align*}
		where $K := \bigcup_{i=1}^nK_i$ and $g_{|K}$ is the restriction of $g$ to $K$.
		\label{prp:1}
	\end{proposition}
	
	The universality property of a kernel helps us in finding the decomposition of matrix $\mathbf{M}$ in the feature space. Following Proposition \eqref{prp:1}, for each row of $\mathbf{M}$, we can always find $h\in\mathbb{H}$ to approximate the row values in a higher dimensional inner product space. We can choose node representations from the disjoint subsets, but note that, each element $h\in\mathbb{H}$ does not have to be in the image of the feature map. 
	
	% \begin{corollary}
	%     Let $0<r\leq \infty$ and $f:(-r, r) \rightarrow \mathbb{R}$ be a $C^{\infty}$-function that can be expanded its Taylor series in $0$, i.e
	%     \begin{align*}
	%         f(x) = \sum_{n=0}^{\infty}a_nx^n && \forall x \in (-r, r)
	%     \end{align*}
	%     Let $\mathsf{X} := \{x \in \mathbb{R}^d: \norm{X}_2 < \sqrt{r}\}$. If we have $a_n >0$ for all $n \geq 0$ then $k(x,y) := f(\langle x, y \rangle)$ defines a universal kernel on every compact subset of $\mathsf{X}$.
	%     \end{corollary}

	Based on the above, we move the inner product from space $\textsf{X}$ to the feature space $\mathbb{H}$, by reformulating Eq. (\ref{eq:objective_without_kernel}) as follows:

	\begin{align}
	&\argmin_{\mathbf{A},\mathbf{B}}\frac{1}{2}\sum_{\boldsymbol{w}\in\mathcal{W}}\sum_{w_l \in \boldsymbol{w}}\sum_{u \in \mathcal{V}}\Big(\mathbf{M}_{w_l,u}^{\textbf{w}} \!\! - \langle \Phi(\mathbf{A}_{w_l,:}),\Phi(\mathbf{B}_{u,:})\rangle \Big)^2\nonumber \\
	=&\argmin_{\mathbf{A},\mathbf{B}}\frac{1}{2}\sum_{\boldsymbol{w}\in\mathcal{W}}\sum_{w_l \in \boldsymbol{w}}\sum_{u \in \mathcal{V}}\Big(\mathbf{M}_{w_l,u}^{\textbf{w}} \!\! - \kappa(\mathbf{A}_{w_l,:},\mathbf{B}_{u,:}) \Big)^2\label{eq:main_obj_kernel}.
	\end{align}
	
	%\noindent where $\mathbf{A}_{w_l}$ and $\mathbf{B}_{u}$ denote the rows of the matrices corresponding to nodes $w_l\in\mathcal{V}$ and $u\in\mathcal{V}$, respectively.
	
	\noindent That way, we obtain a kernelized matrix factorization model for node embeddings based on random walks. For the numerical evaluation of our method, we use the following universal kernels \cite{universal_kernels, kernel_steinwart}:
	
	\begin{align*}
	\kappa_{G}(x,y) & = \exp\left(\frac{-\left\Vert x-y\right\Vert^2}{\sigma^2}\right) && \sigma \in \mathbb{R}\\
	\kappa_{S}(x,y) & = \frac{1}{\left( 1 + \left\Vert x-y \right\Vert^2\right)^{\alpha}} && \alpha \in \mathbb{R}_+
	\end{align*}
	
	\noindent where $\kappa_{G}$ and $\kappa_{S}$ correspond to the \textit{Gaussian} and \textit{Schoenberg} kernels respectively. We will refer to the proposed kernel-based node embeddings methodology as \textsc{KernelNE} (the two different kernels will be denoted by \textsc{Gauss} and \textsc{Sch}).

	\vspace{.2cm}
	\noindent\textbf{Model Optimization.} For the optimization step, we employ \textit{Stochastic Gradient Descent} (SGD) \cite{sgd}. Note that,  Eq. (\ref{eq:main_obj_kernel}) can be divided into two parts with respect to the values of $\mathbf{M}_{v,u}^{\textbf{w}}\in\{0,1\}$. That way, we apply \textit{negative sampling} \cite{word2vec} which is a variant of \textit{noise-contrastive estimation} \cite{nce}, proposed as an alternative to solve the computational problem of hierarchical softmax. For each context node $u^+ \in \mathcal{C}_{\textbf{w}}(w_l)$, we sample $k$ negative instances $u^-$ from the noise distribution $p^-$:
	
	\begin{align*}
	\Big(1 - \kappa(\mathbf{A}_{v,:},\mathbf{B}_{u^+,:})\Big)^2 + \sum_{u^{-} \sim p^-}\Big(\kappa(\mathbf{A}_{v,:},\mathbf{B}_{u^-,:})\Big)^2.
	\end{align*}
	
	\noindent Each sample is generated proportionally to its frequency raised to the power of $0.75$ and the number of negative instances is chosen as $5$. In our experiments, we set the initial learning rate of SGD to $0.025$; then it decreases linearly  according to the number of processed nodes.
	%In addition, we reset it to the minimum value equal to $0.0001$, if it falls below this value.
	The dimension of the embedding vectors is selected as $d=128$ and the window size for the random walks as $\gamma=10$.

	\section{Numerical Tests}
	We evaluate the performance of our approach on the node classification and link prediction tasks. The experiments have been performed on a server with 60Gb RAM. Table \ref{tab:networks} gives the statistics of the network datasets used in the experiments (all the networks are considered as undirected).
	
	%The C++ implementation code of the models are provided for reproducibility purposes of the study in Sec. \ref{sec:introduction}.
	
	\subsection{Baseline Methods}
	We consider five widely used baseline models to compare the performance of our approach. \textsc{DeepWalk} \cite{deepwalk} performs uniform random walks to generate the context of a node; then, the Skip-Gram model is used to learn node representations.
	%The same process is repeated until the desired number of walks is generated.
	\textsc{Node2Vec} \cite{node2vec} combines Skip-Gram with biased random walks, using two extra parameters that control the  walk  to simulate a \textit{BFS} or \textit{DFS} exploration. In the experiments, we set those parameters to $1.0$, the number of walks to $80$ and walk length to $10$. In our approach, we sample context nodes using \textsc{Node2Vec}'s random walk strategy. \textsc{LINE} \cite{line} learns embeddings relying on first-order and second-order proximity information of nodes. \textsc{HOPE} \cite{hope} is a matrix factorization approach aiming at capturing higher-order node similarity patterns based on the \textit{Katz} index. Lastly, \textsc{NetMf} \cite{netmf} targets to factorize the matrix approximated by the pointwise mutual information of center and context pairs. Those methods are compared against the \textsc{KernelNE-Gauss} and \textsc{KernelNE-Sch} models.
	
	%Those methods are compared against the initial model of Eq. \ref{eq:objective_without_kernel} (NE) and the \textsc{KernelNE-Gauss} and \textsc{KernelNE-Sch} models.
	
	\begin{table}[!t]
		\caption{Statistics of networks used in the experiments. $|\mathcal{V}|$: number of nodes, $|\mathcal{E}|$: number of edges, $|\mathcal{K}|$: number of labels and $|\mathcal{C}|$: number of connected components.}
		\label{tab:networks}
		\vspace{-.2cm}
		\renewcommand{\arraystretch}{1.3}
		\resizebox{0.49\textwidth}{!}{%
			\begin{tabular}{rcccccl}
				\hline
				\multicolumn{1}{l}{} & \textbf{$|\mathcal{V}|$} & \textbf{$|\mathcal{E}|$} & \textbf{$|\mathcal{K}|$} & \textbf{$|\mathcal{C}|$} & \textbf{Avg. Degree} &  \textbf{Type} \\\hline
				\textsl{CiteSeer} \cite{harp} & 3,312 & 4,660 & 6 & 438 & 2.814 & Citation \\
				\textsl{Cora} \cite{cora} & 2,708 & 5,278 & 7 & 78 & 3.898  & Citation \\
				\textsl{DBLP} \cite{dblpdataset} & 27,199 & 66,832 & 4 & 2,115 & 4.914  & Co-authorship\\\hline
				%\textsf{PPI} & 3,890 & 38,739 & 50 & 35 & 19.917 & 0.0051 & \\\hline
				\textsl{AstroPh} \cite{astroph} & 17,903 & 19,7031 & - & 1 & 22.010 &  Collaboration \\
				\textsl{HepTh} \cite{astroph} & 8,638 & 24,827 & - & 1 & 5.7483 & Collaboration\\
				\textsl{Facebook} \cite{facebook} & 4,039 & 88,234 & - & 1 & 43.6910 &  Social\\
				\textsl{Gnutella} \cite{astroph} & 8,104 & 26,008 & - & 1 & 6.4186& Peer-to-peer \\\hline
			\end{tabular}%
		}
	\end{table}
	
	% \subsection{Datasets}
	% In order to be consistent in the experiments, we consider each network as undirected and the detailed statistics of the datasets are given in Table \ref{tab:networks}.

	% \begin{itemize}
	% 	\item \textsf{CiteSeer} \cite{harp} is a citation network obtained from the \textit{CiteSeer} library, in which each node corresponds to a paper and the edges indicate  reference relationships among papers. The labels represent the subjects of the paper. 
	
	% 	\item \textsf{Cora} \cite{cora} is another citation network constructed from the publications in the machine learning area; the documents are classified into seven categories. 
	
	% 	\item \textsf{DBLP} \cite{dblpdataset} is a co-authorship graph, where an edge exists between nodes if two authors have co-authored at least one paper. The labels represent the research areas. 
	
	% 	\item Protein-Protein Interactions (\textsf{PPI}) \cite{node2vec} is a graph extracted from the PPI network for Homo Sapiens in which biological states are used as labels of nodes.
	
	% 	\item \textsf{AstroPh} \cite{astroph} is another collaboration network built from the papers submitted to the  \textit{ArXiv} repository for the Astro Physics subject area, from January 1993 to April 2003.
	
	% 	\item \textsf{HepTh} \cite{astroph} network is constructed in a similar way from the papers submitted to \textit{ArXiv} for the \textit{High Energy Physics - Theory} category.
	
	% 	\item \textsf{Gnutella09} 
	
	% 	\item \textsf{Facebook} \cite{facebook} is a social network extracted from a survey conducted via a \textit{Facebook} application.
	% \end{itemize}

	\subsection{Node Classification}
	\noindent \textbf{Experimental set-up.} In the node classification task, we have access to the labels of a certain fraction of nodes in the network (training set), and  our goal is to predict the labels of the remaining nodes (test set). In the experiments, we learn embeddings on varying sizes of training data, ranging from $1\%$ up to $90\%$. The experiments have been carried out by applying an one-vs-rest logistic regression classifier with $L_2$ regularization; the average scores of $50$ experiments are reported.
	
	\vspace{.1cm}
	\noindent \textbf{Experimental results.} Table \ref{tab:classification} shows the Micro-$F_1$ scores for each network. For the \textsl{CiteSeer} network, \textsc{KernelNE} outperforms the baselines for all training sizes. The \textsc{Sch} kernel with $\alpha=1$ gives gain of up to $7.0\%$ against the best baseline model, while \textsc{KernelNE-Gauss} with $\sigma^2=2$ has the best performance for larger training sizes. For the \textsl{Cora} network, the \textsc{Gauss} kernel with  $\sigma^2=2$ also performs quite well especially for small training ratios (gain $0.91\%$ up to $7.60\%$. Lastly, in the \textsl{DBLP} network, we choose $\sigma = 0.3$ for the \textsc{Gauss} kernel, which is the best performing model. The \textsc{Sch} kernel with $\alpha=3.0$, also performs better than the baselines, especially for small training sizes.

	\begin{table}[!t]
		\centering
		\caption{Micro-$F_1$ scores for the node classification task.}
		\label{tab:classification}
		\vspace{-.5cm}
		\subfloat[\textsl{CiteSeer}]{
			\resizebox{0.48\textwidth}{!}{%
				\renewcommand{\arraystretch}{1.3}
				\begin{tabular}{r|cccccccccc}
					\multicolumn{1}{l}{} & \textbf{2\%} & \textbf{4\%} & \textbf{6\%} & \textbf{8\%} & \textbf{10\%} & \textbf{30\%} & \textbf{50\%} & \textbf{70\%} & \textbf{90\%} \\ \hline
					\textsc{DeepWalk} & 0.416 & 0.460 & 0.489 & 0.505 & 0.517 & 0.566 & 0.584 & 0.595 & 0.592 \\
					\textsc{Node2Vec} & 0.450 & 0.491 & 0.517 & 0.530 & 0.541 & 0.585 & 0.597 & 0.601 & 0.599 \\
					\textsc{LINE} & 0.323 & 0.387 & 0.423 & 0.451 & 0.466 & 0.532 & 0.551 & 0.560 & 0.564 \\
					\textsc{HOPE} & 0.196 & 0.205 & 0.210 & 0.204 & 0.219 & 0.256 & 0.277 & 0.299 & 0.320 \\
					\textsc{NetMF} & 0.451 & 0.496 & 0.526 & 0.540 & 0.552 & 0.590 & 0.603 & 0.604 & 0.608 \\\hline
					%\textsc{NE} & 0.479 & 0.510 & 0.529 & 0.547 & 0.557 & 0.596 & 0.612 & 0.620 & 0.623 \\ 
					\textsc{Gauss} & 0.479 & 0.514 & 0.535 & 0.548 & 0.560 & \textbf{0.603} & \textbf{0.615} & \textbf{0.623} & \textbf{0.630} \\
					\textsc{Sch} & \textbf{0.482} & \textbf{0.519} & \textbf{0.538} & \textbf{0.552} & \textbf{0.561} & 0.599 & 0.613 & 0.620 & 0.627 \\ \hline
				\end{tabular}%
			}
			\label{fig:classification_citeseer}
		}
		\vfil
		\vspace{-.2cm}
		\subfloat[\textsl{Cora}]{
			\resizebox{0.48\textwidth}{!}{%
				\renewcommand{\arraystretch}{1.3}
				\begin{tabular}{r|cccccccccc}
					\multicolumn{1}{l}{} & \textbf{2\%} & \textbf{4\%} & \textbf{6\%} & \textbf{8\%} & \textbf{10\%} & \textbf{30\%} & \textbf{50\%} & \textbf{70\%} & \textbf{90\%} \\ \hline
					\textsc{DeepWalk} & 0.621 & 0.689 & 0.715 & 0.732 & 0.747 & 0.802 & 0.819 & 0.826 & 0.833 \\
					\textsc{Node2Vec} & 0.656 & 0.714 & 0.743 & 0.757 & 0.769 & 0.815 & 0.831 & 0.839 & 0.841 \\
					\textsc{LINE} & 0.450 & 0.544 & 0.590 & 0.633 & 0.661 & 0.746 & 0.765 & 0.774 & 0.775 \\
					\textsc{HOPE} & 0.277 & 0.302 & 0.299 & 0.302 & 0.302 & 0.301 & 0.302 & 0.303 & 0.302 \\
					\textsc{NetMF} & 0.636 & 0.716 & 0.748 & 0.767 & 0.773 & \textbf{0.821} & \textbf{0.834} & \textbf{0.841} & \textbf{0.844} \\ \hline
					%\textsc{NE} & \textbf{0.709} & 0.743 & 0.760 & 0.768 & 0.776 & 0.806 & 0.816 & 0.827 & 0.829 \\
					\textsc{Gauss} & \textbf{0.706} & \textbf{0.746} & \textbf{0.761} & \textbf{0.774} & \textbf{0.782} & 0.815 & 0.830 & 0.837 & 0.842 \\
					\textsc{Sch} & 0.693 & 0.733 & 0.753 & 0.761 & 0.769 & 0.799 & 0.810 & 0.819 & 0.824 \\ \hline
				\end{tabular}%
			}
			\label{fig:classification_cora}
		}
		\vfil
		\vspace{-.2cm}
		\subfloat[\textsl{DBLP}]{
			\resizebox{0.48\textwidth}{!}{%
				\renewcommand{\arraystretch}{1.3}
				\begin{tabular}{r|cccccccccc}
					\multicolumn{1}{l}{} & \textbf{2\%} & \textbf{4\%} & \textbf{6\%} & \textbf{8\%} & \textbf{10\%} & \textbf{30\%} & \textbf{50\%} & \textbf{70\%} & \textbf{90\%} \\ \hline
					\multicolumn{1}{r|}{\textsc{DeepWalk}} & 0.545 & 0.585 & 0.600 & 0.608 & 0.613 & 0.626 & 0.628 & 0.628 & 0.633 \\
					\multicolumn{1}{r|}{\textsc{Node2Vec}} & 0.575 & 0.600 & 0.611 & 0.619 & 0.622 & 0.636 & 0.638 & 0.639 & 0.639 \\
					\multicolumn{1}{r|}{\textsc{LINE}} & 0.554 & 0.580 & 0.590 & 0.597 & 0.603 & 0.618 & 0.621 & 0.623 & 0.623 \\
					\multicolumn{1}{r|}{\textsc{HOPE}} & 0.379 & 0.378 & 0.379 & 0.379 & 0.379 & 0.379 & 0.379 & 0.378 & 0.380 \\
					\multicolumn{1}{r|}{\textsc{NetMF}} & 0.577 & 0.589 & 0.596 & 0.601 & 0.605 & 0.617 & 0.620 & 0.623 & 0.623 \\\hline
					%\multicolumn{1}{r|}{\textsc{NE}} & 0.563 & 0.578 & 0.586 & 0.591 & 0.596 & 0.606 & 0.607 & 0.610 & 0.612 \\ 
					\multicolumn{1}{r|}{\textsc{Gauss}} & \textbf{0.611} & \textbf{0.621} & \textbf{0.626} & \textbf{0.628} & \textbf{0.630} & \textbf{0.637} & \textbf{0.641} & \textbf{0.642} & \textbf{0.644} \\
					\multicolumn{1}{r|}{\textsc{Sch}} & 0.610 & 0.616 & 0.622 & 0.624 & 0.625 & 0.633 & 0.636 & 0.637 & 0.638 \\ \hline
				\end{tabular}%
			}
			\label{fig:classification_dblp}
		}
	\end{table}

	\subsection{Link Prediction}
	\noindent \textbf{Experimental set-up.} For the link prediction task, we remove  half of the edges of the network in order to obtain positive samples for the test set; the same number of node pairs, not existing in the initial graph, are added to the test set. We then learn node embedding using the residual network; the feature vector of an edge $(v,u)$ is formed with the operation $|x^v_i-y^u_i|^2$ for each coordinate $i$ of the embedding vectors $\boldsymbol{x}$ and $\boldsymbol{y}$ corresponding to nodes $v$ and $u$. In the experiments, we use logistic regression with $L_2$ regularization.
	
	\begin{table}[!t]
		\renewcommand{\arraystretch}{1.3}
		\caption{Area Under Curve (AUC) scores for the link prediction task.}
		\label{tab:edge_prediction}
		\vspace{-.2cm}
		\centering
		\resizebox{0.48\textwidth}{!}{%
			\begin{tabular}{r|ccccc|cc}
				& \textsc{DeepWalk} & \textsc{Node2Vec} & \textsc{LINE} & $\textsc{HOPE}$ & \textsc{NetMF}  & $\textsc{Gauss}$ & $\textsc{Sch}$ \\\hline
				\textsl{CiteSecer} & 0.837 & 0.762 & 0.557 & 0.756 & 0.742 &\textbf{0.886} & 0.875 \\
				\textsl{Cora} & 0.778 & 0.724 & 0.554 & 0.728 & 0.755  & \textbf{0.819} & 0.814 \\
				\textsl{DBLP} & 0.944 & 0.905 & 0.590 & 0.930 & 0.930  & \textbf{0.963} & 0.958 \\
				\textsl{AstroPh} & 0.960 & 0.935 & 0.679 & 0.967 & 0.897 & \textbf{0.978} & 0.970 \\
				\textsl{HepTh} & 0.897 & 0.830 & 0.633 & 0.875 & 0.882 & \textbf{0.920} & 0.915 \\
				\textsl{Facebook} & 0.983 & \textbf{0.988} & 0.696 & 0.980 & 0.987 & 0.987 & 0.987 \\
				\textsl{Gnutella} & 0.680 & 0.498 & 0.702 & 0.599 & 0.651 & \textbf{0.766} & 0.677\\ \hline
			\end{tabular}%
		}
	\end{table}
	
	\vspace{.1cm}
	\noindent \textbf{Experimental results.} Table \ref{tab:edge_prediction} shows the \textit{area under curve} (AUC) scores for the link prediction task. We choose kernel parameters as $\sigma=0.3$ and $\alpha=2$ except the \textsl{Gnutella} network in which $\sigma=3.0$. In all cases, the largest connected component of the datasets is used. As we can observe, in almost all cases, the proposed kernel-based models outperform the baselines. The only exception is the \textsl{Facebook} dataset, where \textsc{Node2Vec} is just slightly better than \textsc{KernelNE}. 
	
	%The only exception is the \textsl{Facebook} dataset, where NE, the basic matrix factorization model of Eq. \eqref{eq:objective_without_kernel}, has the same performance as \textsc{Node2Vec}. 
	
	\subsection{Parameter Sensitivity}
	% Here we examine how the performance of the kernel  methods is affected based on the choice of kernel parameters.    
	
\begin{figure}[!t]
		\centering
		\includegraphics[width=0.45\textwidth]{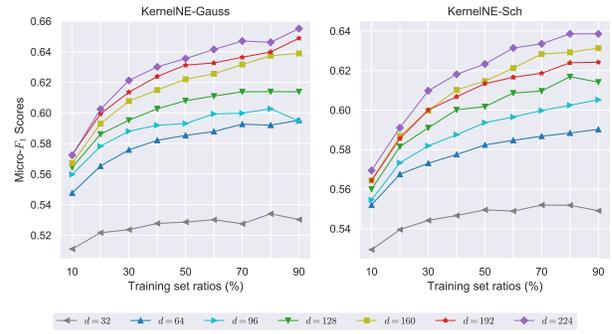}
		\vspace{-.2cm}
		\caption{Influence of the dimension size $d$ on the \textsl{CiteSeer} network.}
		\label{fig:dimension_size}
	\end{figure}

	\vspace{.1cm}
	\noindent \textbf{The effect of dimension size.} Figure \ref{fig:dimension_size} shows the Micro-$F_1$ scores of the proposed models for varying embedding dimension sizes, ranging from $d=32$ up to $d=224$. As it can be seen,  both of the kernel instances have the same  tendency, where the performance increases proportionally to the size of the embedding vectors.

	\begin{figure}[!t]
		\centering
		\subfloat{\includegraphics[width=0.225\textwidth]{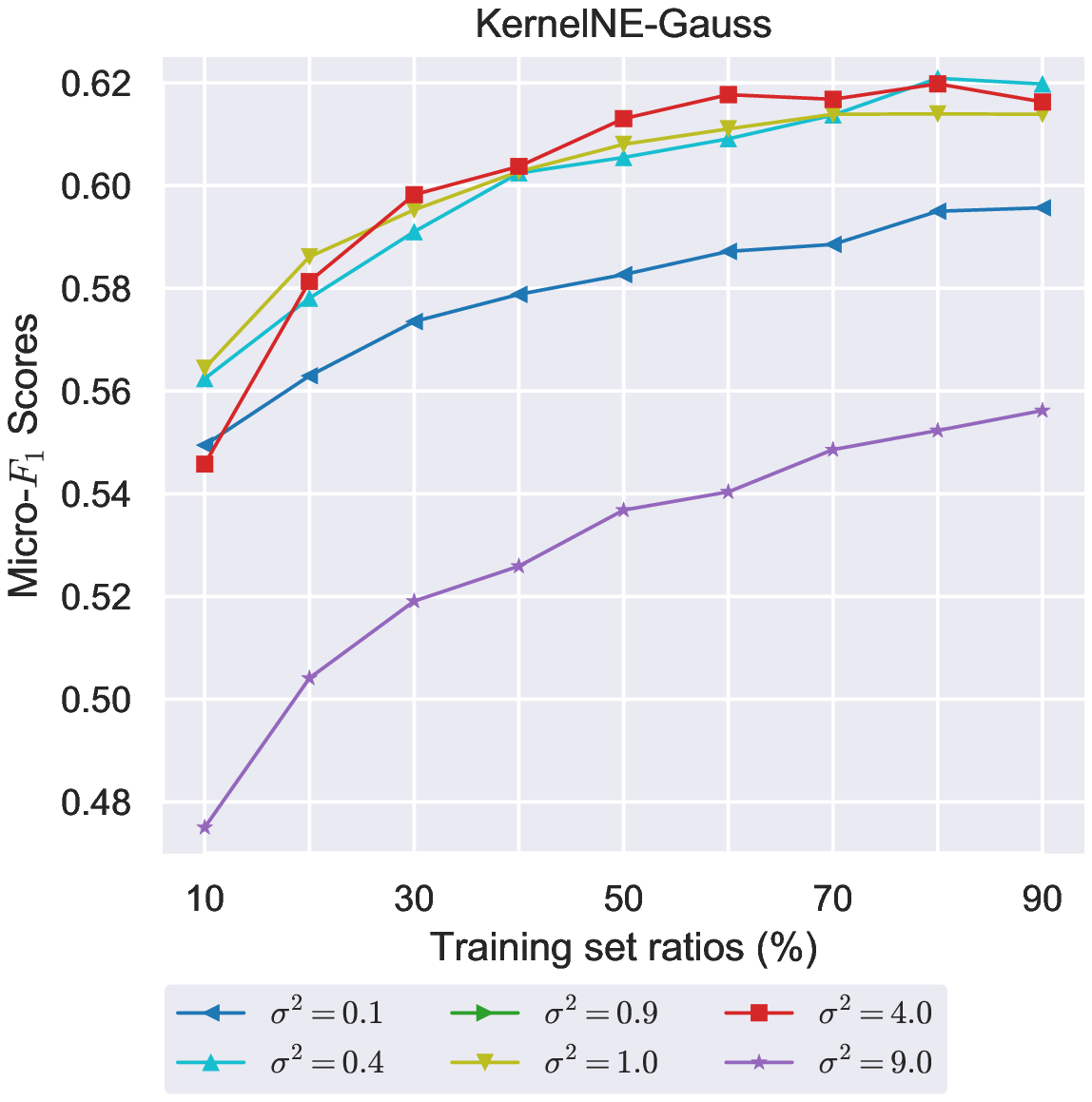}%
		}
		\hfil
		\subfloat{\includegraphics[width=0.225\textwidth]{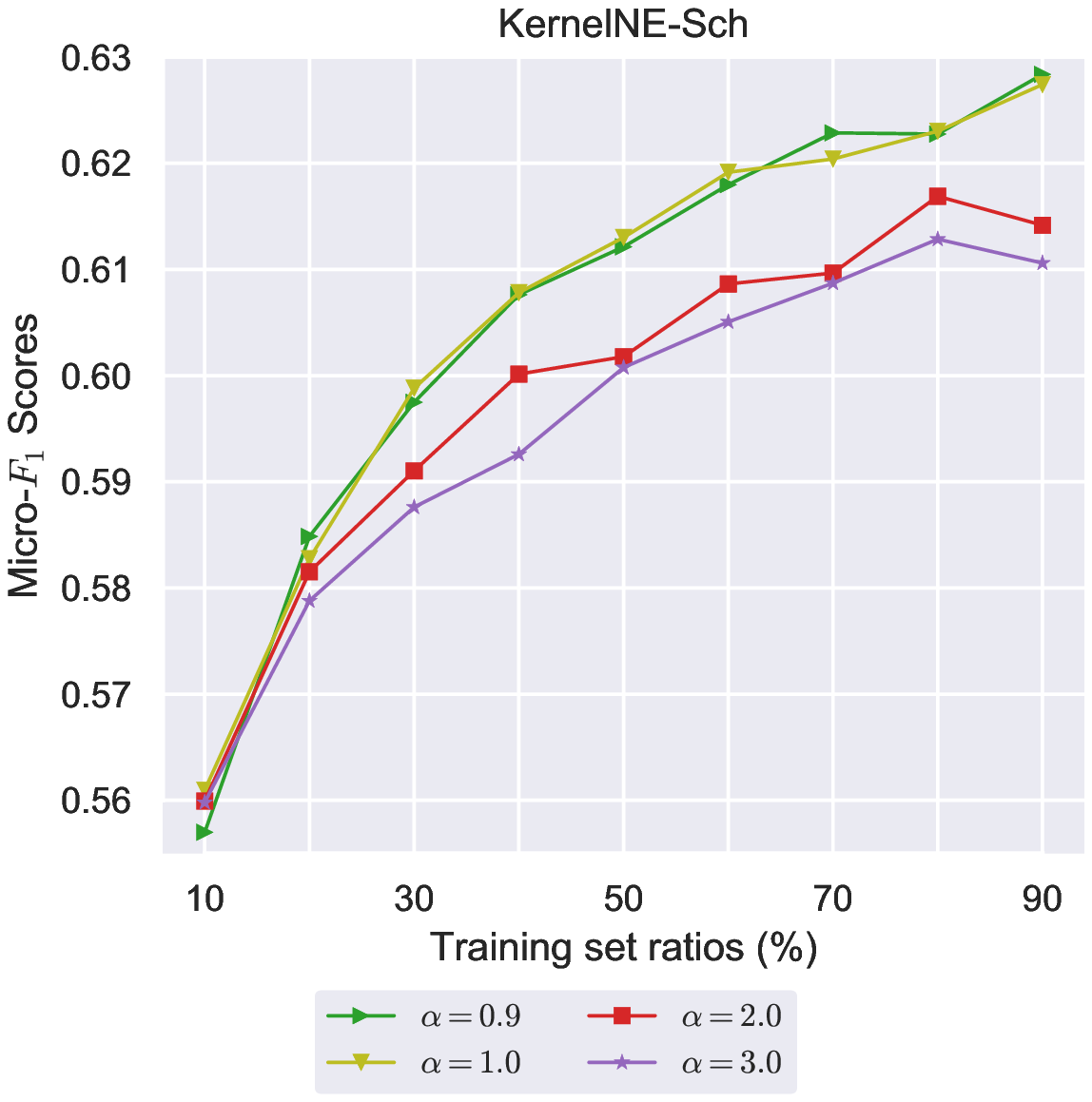}%
		}
		\vspace{-.2cm}
		\caption{Influence of kernel parameters on the \textsl{CiteSeer} network.}
		\label{fig:kernel_params}
	\end{figure}
	
	\vspace{.1cm}
	\noindent \textbf{The effect of kernel parameters.} In Figure \ref{fig:kernel_params}, we study the behaviour of kernel functions with respect to the chosen parameters. The \textsc{Gauss} kernel shows comparable results for values of $\sigma^2$  between $0.4$ and $4.0$. In addition, we observed that its performance is limited for very big or very small values of this parameter. The \textsc{Sch} kernel also behaves similarly.We have reached the highest score with parameter values around $\alpha=1.0$. Lastly, we observed poor performance for very small values of  $\alpha$, which are not included in the figure.

	\section{Conclusion}
	%In the paper
	We have introduced the \textsc{KernelNE} model for learning node embeddings. We interpret our random-walk based method under a weighted matrix factorization framework, which is then generalized to kernel functions. The numerical evaluation showed that the proposed kernel-based models substantially outperform baseline NRL methods in both node classification and link prediction tasks. An interesting future research direction concerns the extension of the proposed methodology to the multiple kernel learning framework \cite{BachMKL}.


	% % use section* for acknowledgment
	% \section*{Acknowledgment}


	
	
	%\bibliographystyle{IEEEtran}
	
	
	% \end{thebibliography}

	\bibliographystyle{IEEEtran}
	\balance
	\bibliography{main}
	
\end{document}